\documentclass[aps,12pt,superscriptaddress,preprint]{revtex4}
\usepackage{graphicx}
\usepackage{amsmath}
\usepackage{hyperref}
\usepackage{bm}
\usepackage{CJK}
\usepackage[dvips]{color}

\graphicspath{{figs/},
      }

\newcommand{\degr}{$^{\circ}$}

\newcommand{\fzd} {Institute of Ion Beam Physics and Materials Research, Forschungszentrum Dresden-Rossendorf, P.O. Box 510119, 01314 Dresden, Germany}

\newcommand{\tud} {Institut f\"{u}r Theoretische Physik, Technische Universit\"{a}t Dresden, 01062 Dresden, Germany}

\newcommand{\innovavent} {INNOVAVENT GmbH, Bertha-von-Suttner-Str. 5, 37085 G\"{o}ttingen, Germany}

\begin{document}
\begin{CJK*}{GB}{gbsn}
\title{Hysteresis in the magneto-transport of Manganese-doped Germanium: evidence for carrier-mediated ferromagnetism}
\date{\today}

\author{Shengqiang~Zhou (ÖÜÉúÇ¿)}
\email[Electronic address: ]{S.Zhou@fzd.de} \affiliation{\fzd}
\author{Danilo~B\"{u}rger}
\author{Arndt~M\"{u}cklich}
\author{Christine~Baumgart}
\author{Wolfgang~Skorupa}
\affiliation{\fzd}
\author{Carsten~Timm}
\affiliation{\tud}
\author{Peter~Oesterlin}
\affiliation{\innovavent}
\author{Manfred~Helm}
\affiliation{\fzd}
\author{Heidemarie~Schmidt}
\affiliation{\fzd}

\begin{abstract}
We report the fabrication of Ge:Mn ferromagnetic semiconductors by Mn-ion implantation into Ge followed by pulsed laser annealing. Benefiting
from the short time annealing, the hole concentration in Mn-implanted Ge has been increased by two orders of magnitude from 10$^{18}$ to over
10$^{20}$ cm$^{-3}$. Likely due to the high hole concentration, we observe that the longitudinal and Hall resistances exhibit the same hysteresis as the
magnetization, which is usually considered as a sign of carrier-mediated ferromagnetism.

\end{abstract}
\maketitle
\end{CJK*}
\section{Introduction}
Diluted ferromagnetic semiconductors (FMS) are considered to be promising materials for spintronic applications. FMS exhibit strong
magneto-transport effects, namely negative magnetoresistance (MR) and the anomalous Hall effect (AHE).\cite{PhysRevLett.68.2664,PhysRevB.63.085201} GaAs:Mn has been intensively investigated during the last 20 years and is considered as the
prototype FMS. Practically, it is desirable to have a FMS compatible with silicon technology. Manganese-doped germanium prepared using
low-temperature molecular-beam epitaxy (LT-MBE) proves to be most promising due to the effective suppression of precipitate formation.\cite{Park01252002,jamet06,bougeard:237202,li:201201} Two conclusions have been well established: (i) Mn segregation is unavoidable even at
growth temperatures below 70 {\degr}C.\cite{Park01252002,jamet06,bougeard:237202,li:201201,PhysRevB.76.205306} (ii) Mn-rich precipitates,
mainly Mn$_{5}$Ge$_{3}$ and Mn$_{11}$Ge$_{8}$, form inside the Mn-diluted Ge matrix when the growth temperature is higher than 70 {\degr}C
\cite{li:152507,ahlers:214411} or during ion implantation at elevated temperatures \cite{zhou_Mn5Ge3}. On the other hand, various experiments
evidence that diluted Mn ions are in the 2+ state \cite{PhysRevLett.94.147202,picozzi:062501,gambardella:125211} and act as double acceptors in
Ge. A substitutional occupation of Fe, Cu, and Ag in Ge was confirmed recently. \cite{PhysRevLett.102.065502} Thus, Mn-rich regions are embedded
inside the Ge matrix together with substitutional Mn ions. The appearance of Mn-rich regions can be demonstrated by transmission electron
microscopy and by zero-field-cooling/field-cooling (ZFC/FC) magnetization measurements. \cite{jamet06,li:152507,bougeard:237202}

We notice that pronounced MR and AHE in the Ge:Mn system have been reported.\cite{Park01252002,PhysRevLett.91.177203,jamet06,li:201201,zeng:066101,deng:062513} However, there are fundamental open questions. First of
all, the correlation between magnetization, MR, and AHE, which is a hallmark of the GaAs:Mn FMS,\cite{PhysRevLett.68.2664,PhysRevB.63.085201}
has not been proven for Ge:Mn so far. For isotropic and anisotropic magnetic samples the MR and AHE curves measured in van der Pauw geometry
always reflect the field dependence of magnetization perpendicular to the sample surface. For example, GaAs:Mn grown on GaAs(001) has its hard
axis of magnetization perpendicular to the sample surface. As expected the corresponding AHE curve does also not show an open
hysteresis loop and is perfectly correlated with the hard axis magnetization perpendicular to the sample surface.\cite{ohnoGaMnAs} The reported
MR and AHE in Ge:Mn are likely caused by superparamagnetic Mn ions or precipitates \cite{riss:241202} or by a two-band-like conduction.\cite{ZhouAHE} Note that only paramagnetic coupling between Mn impurities in Ge:Mn was revealed by X-ray magnetic circular dichroism
\cite{gambardella:125211}. Secondly, measurements of the resistivity versus temperature often reveal a small activation energy of several meV at
low temperatures, \cite{Park01252002,PhysRevLett.91.177203,PhysRevB.72.165203,riss:241202} which is much smaller than the thermal ionization
energy of Mn in Ge. \cite{PhysRev.100.659} Thirdly, the AHE curves measured above 10 K usually reveal a reversal in slope.
\cite{PhysRevLett.91.177203,jamet06,PhysRevB.72.195205,zeng:066101} On the other hand, below \mbox{10 K} the AHE has been reported to be absent.
\cite{riss:241202}

We believe that the origin of these puzzling observations lies in the less effective substitution of Mn at Ge sites, which results in too low a
hole concentration, making carrier-mediated ferromagnetism impossible. In previous work, the ferromagnetism in Ge:Mn has been non-quantitatively but plausibly
explained by the formation of bound magnetic polarons (BMP). \cite{Park01252002,PhysRevB.72.195205,li:152507,bougeard:237202} According to the
model of Kaminski and Das Sarma \cite{PhysRevB.68.235210} for GaAs:Mn with 8.5\% Mn (Ref. \onlinecite{PhysRevB.70.193203}) the critical hole
concentration for percolation at 10 K amounts to \mbox{3$\times$10$^{19}$ cm$^{-3}$}. The hole concentrations realized in Ge:Mn grown by LT-MBE
\cite{Park01252002,li:201201,PhysRevLett.91.177203,jamet06,Ahlers2006422,PhysRevB.72.165203} are mostly well below the threshold value of
\mbox{3$\times$10$^{19}$ cm$^{-3}$}, which indicates the possible unsuitability of the LT-MBE approach to achieve a large hole concentration in Ge:Mn.

In this article we show that the hole concentration in Mn-implanted Ge can be strongly increased by short-time annealing. The largest hole
concentration achieved is around \mbox{$2.1\times10^{20}$ cm$^{-3}$} in samples after pulsed laser annealing (PLA), showing negative MR and AHE
with the same hysteresis as the magnetization at low temperatures.

\section{Experimental}
Nearly intrinsic, n-type Ge(001) wafers were implanted with Mn ions. The implantation energy and fluence were 100 keV, 30 keV and
5$\times10^{16}$ cm$^{-2}$, 1$\times10^{16}$ cm$^{-2}$, respectively, resulting in a box-like distribution of Mn ions with concentration around
10 \% over a depth of 100 nm. During implantation the wafers were flow-cooled with liquid nitrogen to avoid the formation of any Mn-rich
secondary phase. Pulsed laser annealing was performed using a laser ASAMA 80-8 and an optical system VOLCANO from INNOVAVENT GmbH. The pulse
duration was 300 ns at a wavelength of 515 nm. The introduced energy density was 1.5 J/cm$^2$. During annealing, the sample surface was scanned
by a laser beam which was focused to a 2 mm$\times$40 $\mu$m large stripe with a frequency of 50 kHz, instead of a large, symmetric
laser spot of \mbox{5$\times$5 mm$^2$} used for GaAs:Mn. \cite{danilo_PRB,scarpulla:073913} Structural analysis were performed using secondary ion mass
spectrometry (SIMS) and transmission electron microscopy (TEM, FEI Titan). Magnetization
properties were investigated with a superconducting quantum interference device (SQUID, Quantum Design MPMS) magnetometer. Magneto-transport
was measured with a magnetic field applied perpendicularly to the film plane in van der Pauw geometry.

\section{Results and Discussion}

\subsection{Structural properties}

\begin{figure} \center
\includegraphics[scale=0.45]{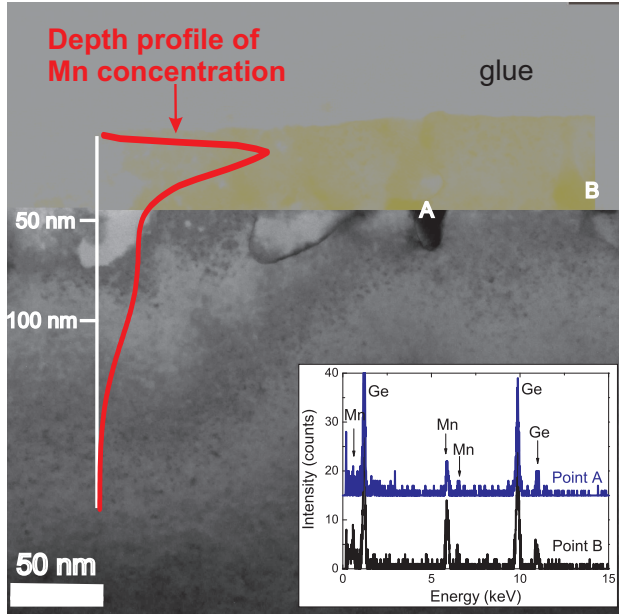}
\caption{Cross-sectional TEM image recorded on PLA annealed Mn-implanted Ge. The image reveals Mn-rich tadpole-like regions,
labeled A and B. Relative compositions of region A and B are given in units of atomic percent in the corresponding EDS spectra (see inset).
The vertical scale bar indicates the distance from the sample surface.
The depth of the implanted and molten layer amounts to ca. 100 nm. There is no structural change between the
implanted/unimplanted layers after PLA. The depth profile of the Mn concentration obtained by SIMS is shown as the red curve.}\label{fig:EDX}
\end{figure}

Fig. 1 shows the cross-sectional TEM image. In a
ca. 60 nm thick top layer there are some tadpole-like regions with a size of 10--20 nm.
Relative composition measurements by means of energy dispersive spectroscopy (EDS)
reveal that those tadpole-like regions are Mn-rich (inset of Fig. \ref{fig:EDX}).
For instance, there are around 19 at.\% and 26 at.\% Mn in
region A and B, respectively. This large Mn concentration does not correspond to any known Ge-Mn
phase \cite{jamet06}. On the other hand, the Mn-rich regions in the top layer are not percolating.
Using high-resolution TEM, we found that the areas between the Mn-rich structures
and the regions in the depth of 60--100 nm are coherent with the single crystalline Ge matrix. Using SIMS (shown as the read curve
along the depth scale in Fig. \ref{fig:EDX}),
we observed an accumulation of Mn in the ca. 40 nm thick top layer after PLA.
In the depth between 50--100 nm, Mn ions are uniformly distributed. We do not have an accurate absolute calibration, but the measurement is
consistent with a Mn concentration of a few percent in this region. Therefore, by structural analysis we can conclude that after PLA the near-surface
layer of Mn-implanted Ge contains Mn-rich tadpole-like regions, while most of the Mn is incorporated in a single crystalline Ge:Mn matrix.

\subsection{Magnetism and Magneto-transport}
\begin{figure} \center
\includegraphics[scale=0.8]{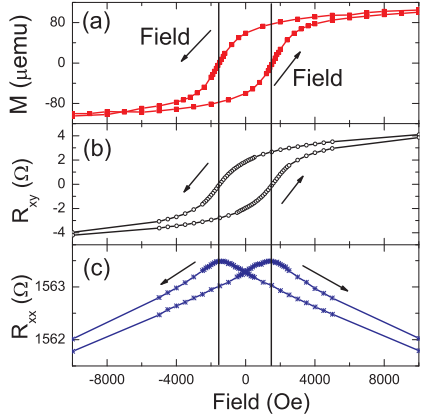}
\caption{The field dependent (a) magnetization (\emph{M}), (b) Hall resistance (\emph{R}$_{xy}$) and (c) longitudinal resistance
(\emph{R}$_{xx}$) at \mbox{5 K}. Hysteresis appears for both \emph{R}$_{xx}$ and \emph{R}$_{xy}$ curves, with the same coercivity as for
\emph{M}. The longitudinal resistivity is around 0.015
$\Omega$-cm if assuming a thickness of 100 nm.}\label{fig:MH_AHE_MR}
\end{figure}

Fig. \ref{fig:MH_AHE_MR} shows the comparison of the field-dependent magnetization, Hall resistance, and longitudinal resistance at 5 K measured
 on the Ge:Mn sample after PLA, and represents a central result of the current
paper. The involvement of holes in the ferromagnetism is unambiguously confirmed by the appearance of the AHE, and, especially, by the clear
hysteresis loop in the Hall resistance. Note that the presence of ferromagnetic characteristics in SQUID measurements alone could also be caused
by ferromagnetic precipitates, \emph{e.g.}, Mn$_5$Ge$_3$, in addition to or instead of a FMS. At low temperature, hysteresis appears in both the
Hall and longitudinal resistance curves, with the same coercive field as in the magnetization. Such a correlation between AHE, MR, and
magnetization is usually considered as the signature of FMS, \cite{PhysRevLett.68.2664,PhysRevB.63.085201} where the same set of holes
contribute to ferromagnetism and transport \cite{dietl00}. This also has been observed in compensated GaAs:Mn \cite{PhysRevB.70.193203} and
insulating GaP:Mn. \cite{Scarpulla05} However, to our knowledge, the correlation between AHE, MR, and magnetization has never been observed for
other Ge:Mn samples.

\begin{figure} \center
\includegraphics[scale=0.7]{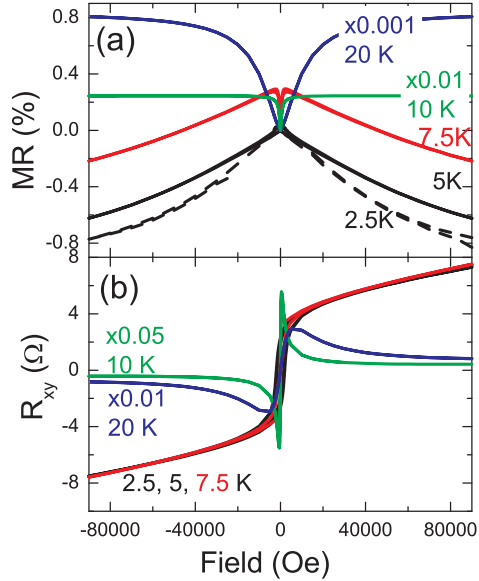}
\caption{(a) Magnetoresistance and (b) Hall resistance measured at different temperatures (for clarity, the curves at 10 K and 20 K are
multiplied by the factors indicated in the figure). The cusp-like features in $R_{xy}$ and electron-like high-field slope at low temperatures
are due to two-band-like conduction \cite{ZhouAHE}.}\label{fig:fig2ab}
\end{figure}

Here we point out that the hysteretic MR and AHE behavior only persists up to around 7.5 K as shown in \mbox{Figs. \ref{fig:fig2ab}(a) and (b)}. By a
linear fitting of the Hall curves at large fields [Fig. \ref{fig:fig2ab}(b)], we can calculate the hole concentration to be 2.1$\times$10$^{20}$
cm$^{-3}$, which does not change between 2.5 and 7.5 K. However, the hole concentration could be largely underestimated due to the influence of
the AHE, as in the case of the GaAs:Mn system. \cite{Omiya2000976} Nevertheless, if we compare the hole concentration with the implantation
fluence, we find that approximately 3.5\% of the Mn atoms result in electrically active holes by neglecting any compensation effect from
interstitial Mn. Above 7.5 K the sample only shows positive MR and AHE without hysteresis, similar as reported in the literature.
\cite{Park01252002,PhysRevLett.91.177203,jamet06,PhysRevB.72.195205,riss:241202,zeng:066101} In Ref. \onlinecite{jamet06}, an extensive
discussion is given to understand the large positive MR in Ge:Mn. We proposed that a two-band-like conduction could induce the large positive MR
and anomalous Hall resistance. \cite{PhysRevB.61.9621,ZhouAHE}

\begin{figure} \center
\includegraphics[scale=0.8]{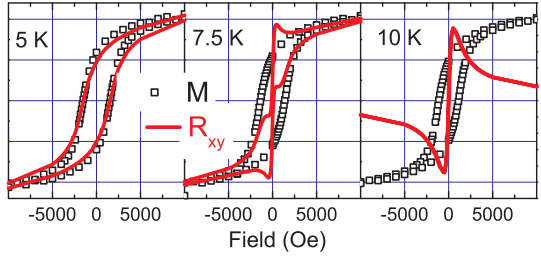}
\caption{Comparison between magnetization (\emph{M}) and Hall resistance (\emph{R}$_{xy}$) curves with the magnetic field from -10000
to 10000 Oe. The signal is scaled for a better visibility. At 5 K, both \emph{M} and \emph{R}$_{xy}$ curves overlap with each other, while above 7.5 K
there is no correlation between \emph{M} and \emph{R}$_{xy}$.}\label{fig:fig2c}
\end{figure}

Above 7.5 K, the carriers contributing mainly to
the electrical transport are not involved in carrier-induced ferromagnetism. Thus, the Hall effect behaves differently from the magnetization
which exhibits hysteresis. As shown in Figs. \ref{fig:fig2c}, at 5 K both \emph{M} and \emph{R}$_{xy}$ curves overlap with each other, while above 7.5 K
there is no correlation between \emph{M} and \emph{R}$_{xy}$.

\begin{figure} \center
\includegraphics[scale=0.7]{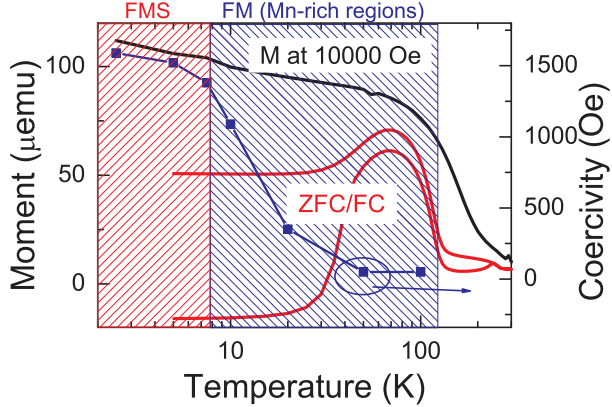}
\caption{Temperature-dependent magnetization at 10000 Oe and the coercivity together with the ZFC/FC magnetization curves at 50
Oe. It can be viewed as a simple magnetic phase diagram of the sample.}\label{fig:fig2d}
\end{figure}

Fig. \ref{fig:fig2d} shows the
temperature-dependent magnetization at 10000 Oe and the coercive
field, as well as the ZFC/FC curves at 50 Oe. We notice three distinct temperature regimes: (i) a weak temperature dependence of the coercive
field below \mbox{10 K}, (ii) a large drop of coercive field above 10 K, and (iii) a large drop of magnetization (in a field of 10000 Oe) at around 100
K. Together with the ZFC/FC curves, we can infer the following: Below \mbox{10 K}, magnetic percolation occurs through the whole sample. Between 10 and
60 K, individual Mn-rich regions are spontaneously magnetized, and by applying a large field, the magnetization of the entire sample decreases
only slightly with increasing temperature. Above the peak in the ZFC/FC curves, the ferromagnetic exchange collapses in most of the Mn-rich
regions and the magnetization drops quickly with increasing temperature. We would like to point out, however, that concerning its magnetization
properties the sample is still ferromagnetic above 100 K, as reported in the literature.
\cite{jamet06,PhysRevB.72.195205,bougeard:237202,li:201201,jaeger:045330}

\begin{figure} \center
\includegraphics[scale=0.8]{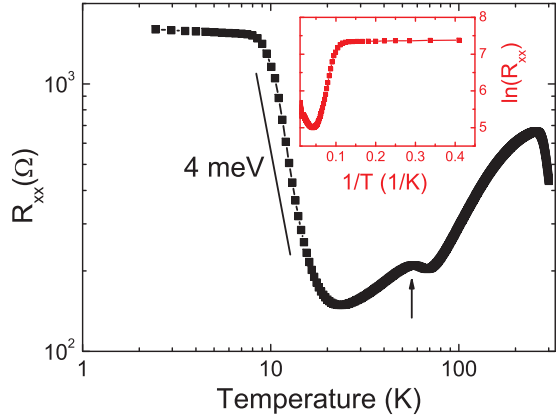}
\caption{Temperature-dependent longitudinal resistance (\emph{R}$_{xx}$) at zero field. The hump around 60--70 K indicated by the arrow
coincides with the peak in the ZFC curves, which is related to Mn-rich regions. The inset shows ln(R$_{xx}$) vs. reciprocal temperature.}
\label{fig:RT}
\end{figure}

The most noticeable feature is the ZFC peak in the temperature range from 60 to 70 K. The peak position does \emph{not} change upon increasing
the field up to 500 Oe (not shown). For normal magnetic nanoparticles the corresponding ZFC peak moves to lower temperatures when applying a
larger magnetic field. \cite{zhou07CoNi} Therefore, the Mn-rich regions are different from crystalline secondary phases. Such Mn-rich regions in
Ge were also observed by other groups, \cite{Park01252002,bougeard:237202,PhysRevB.72.195205} but the magnetic-field-independent ZFC peaks were
found at a much lower temperature around 18 K. The ferromagnetism in those regions is believed to be hole-mediated.
\cite{bougeard:237202,PhysRevB.72.195205} The ZFC/FC peaks in our samples occur at higher temperatures, which indicates a stronger
ferromagnetic coupling inside the Mn-rich regions. According to the model of Kaminski and Das Sarma, \cite{PhysRevB.68.235210} the ferromagnetic
transition in these regions leads to an immediate decrease in the hopping energy which gives rise to a non-monotonic temperature dependence of
the sample resistance. A resistance peak at around \emph{T}$_C$ of Mn-rich regions has been observed experimentally by Li \emph{et al.}
\cite{PhysRevB.72.195205} and is in fact also visible in the temperature-dependent resistance as shown in Fig. \ref{fig:RT}. At low temperature,
there are two regions of fundamentally different behavior: Below \mbox{10 K}, the resistance is nearly constant, \emph{i.e.}, quasi metallic. From 10
to 20 K the resistance decreases steeply with an activation energy of 4 meV. Such a temperature dependence appears to be a rather universal
feature of Mn-doped Ge, \cite{Park01252002,PhysRevLett.91.177203,PhysRevB.72.165203,riss:241202} and is attributed to impurity band conduction.
\cite{PhysRevB.72.165203,jamet06,riss:241202} We will discuss below that this activation energy of 4 meV is related with the
temperature-dependent occupation of the ground and first excited state of the Mn double acceptor which form impurity bands in Ge:Mn.

\subsection{Theoretical picture of Ge:Mn}

\begin{figure} \center
\includegraphics[scale=0.45]{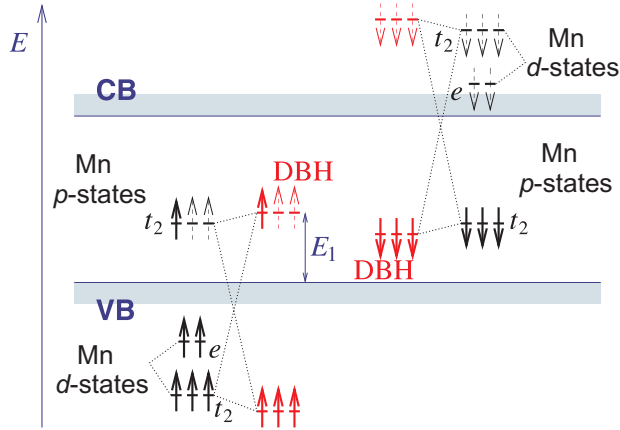}
\caption{Sketch of single-particle states (not to scale) introduced by Mn double acceptors in Ge before (black) and after (red) hybridization,
see text. The spin of the lower lying Mn \textit{d}-shells is assumed to point upward. Spin-orbit coupling is ignored. Bold (dashed) arrows
indicate occupied (unoccupied) states. The valence band (VB) and conduction band (CB) edges are indicated. }\label{fig:hybrid_Ge}
\end{figure}

In analogy to substitutional Mn in GaAs:Mn, \cite{PhysRevB.68.075202} we discuss the microscopic picture of substitutional Mn in Ge. The three Mn \emph{p}-orbitals form
the bound acceptor states of Mn and are subject to a crystal field of tetrahedral symmetry. They belong to the three-dimensional $t_d$ ($\Gamma_5$)
representation of the tetrahedral group $T_d$, which can be interpreted as a state with orbital angular momentum $l=1$. Together with the spin
$s=1/2$, one finds that the acceptor state is sixfold degenerate. In the same $T_d$ crystal field, the five Mn \emph{d}-orbitals split into
three orbitals with $t_2$ character and two with $e$ character. In accordance with Hund's rules for small crystal fields, all five are singly
occupied and the electron spins are aligned. Let us assume that the \emph{d}-shell spin is pointing upward. The five Mn \emph{d}-orbitals with
the \emph{d}-shell spin pointing downward are unoccupied. The separation between the spin upward and spin downward Mn \emph{d}-orbitals (Fig.
\ref{fig:hybrid_Ge}) is mostly due to the on-site Coulomb repulsion. The $t_2$-type spin-up \emph{d}-orbitals hybridize with the $t_2$-type
spin-up acceptor states. This leads to level repulsion, which shifts the spin-up acceptor states ("dangling-bond hybrid", DBH) upward in energy,
as shown in Fig. \ref{fig:hybrid_Ge}. Analogously, level repulsion between the unoccupied spin-down \emph{d}-orbitals and the spin-down acceptor
states shifts the latter downward in energy. Mn is a double acceptor. The hole binding energy is found to be 0.16 eV from the valence band and
0.37 eV from the conduction band for the neutral and singly charged acceptor, respectively. \cite{PhysRev.100.659} In Fig. \ref{fig:hybrid_Ge}
the neutral acceptor is represented as two bound holes going into the single-particle states of highest electronic energy. Consequently, two
spin-up electrons are missing and the total spin is reduced from what one would get from the Mn \emph{p}-orbitals and \emph{d}-orbitals alone. In this sense, the
exchange interaction between the Mn \emph{d}-states and acceptor \emph{p}-states is antiferromagnetic.

To obtain the full picture, many-particle effects and spin-orbit coupling have to be taken into account. In the neutral Mn double acceptor with
two bound holes, \emph{i.e.}, the $d^5+2\mathrm{h}$ complex likely relevant here, the electrons should combine to give a total angular momentum
of magnitude $j=2$ according to Hund's rules. Due to the antiferromagnetic interaction with the Mn \emph{d}-shell spin $S=5/2$ we obtain a total
angular momentum $\mathbf{F}=\mathbf{j}+\mathbf{S}$ with $F=1/2$ in the ground state and $F=3/2$ in the first excited state. The splitting
between the ground-state doublet and the excited state quartet has not been determined so far. We attribute the thermal activation energy of 4
meV from 10 to 20 K as shown in Fig. \ref{fig:RT} to the transition between the total angular momentum states $F=1/2$ and $F=3/2$ of neutral Mn
double acceptors. The exchange interaction can be written as $
    \epsilon\textbf{j}\cdot\textbf{S}=\frac{\epsilon}{2}[F(F+1)-j(j+1)-S(S+1)].
$ Consequently, the splitting between the ground state ($j=2$, $S=5/2$ and $F=1/2$) and the first excited state ($j=2$, $S=5/2$ and $F=3/2$) is
$3\epsilon/2\sim$4 meV and the exchange constant $\epsilon$ is 2.7 meV. This is comparable to $\epsilon$ in GaAs:Mn amounting to 2.5--5.5 meV.
\cite{dietl00} Furthermore, independently of the third Hund coupling compared to $\epsilon$, the calculated spin polarization of the
holes in the ground state amounts to -0.33. If the third Hund coupling is small compared to $\epsilon$, as is likely the case, the calculated
spin polarization in the first excited state amounts to -0.44. Due to the increased spin polarization in the first excited state compared to the
ground state, the hole mobility in the first excited state is likely also higher than in the ground state due to the suppression of spin-flip scattering.
We therefore propose that the thermally activated behavior seen in \mbox{Fig.
\ref{fig:RT}} around 10--20 K is due to the first excited state becoming accessible, which opens another conducting channel with a larger
mobility. However, a definite explanation for the resistance drop is not possible at present due to the strong anomalous Hall effect,
which inhibits a complete determination of carrier concentration and mobility as a function of temperature.

Note that the microscopic picture of an antiferromagnetic coupling between holes and Mn ions is consistent with the macroscopic picture of the
BMP theory proposed by Kaminski and Das Sarma. \cite{PhysRevB.68.235210} When the temperature is low, the ferromagnetic transition will first
occur locally in the regions with larger carrier concentration. As the temperature decreases, these finite-size regions, which have random
shapes and positions, grow and merge, and finally percolation over the whole sample occurs. Therefore, the Ge:Mn FMS material will be rather
described as a combination of Mn-rich and Mn-poor regions with diluted Mn ions. The substitutional Mn ions introduce holes, acting as seeds for
BMPs. The number and size of BMPs increase with the hole concentration and with decreasing temperature, respectively. Within such a picture we
can understand the investigated Ge:Mn sample and also shed light on the absence of hysteretic MR and AHE in the literature. The hole
concentration is the critical parameter for FMS behavior, as seen for GaAs:Mn. \cite{PhysRevB.70.193203} A large enough hole concentration
(2.1$\times$10$^{20}$ cm$^{-3}$) gives rise to the carrier-mediated ferromagnetism throughout the Ge:Mn sample presented here. In addition to
the sample discussed here, we also investigated Ge:Mn samples with lower hole concentration. In those samples we observed ferromagnetism up to
around 100 K, but no hysteretic magneto-transport down to 5 K. \cite{ZhouMnGe} This appears similar to Refs.
\onlinecite{jamet06,PhysRevB.72.195205,bougeard:237202,li:201201,jaeger:045330}, where a ferromagnetic phase with a high \emph{T}$_C$ was
reported due to the Mn-rich regions, but the hysteretic magneto-transport was absent, most likely due to the low hole concentration.
\\

\section{Conclusion}
In conclusion, using the approach of Mn-ion implantation into Ge and subsequent pulsed laser annealing, we have been able to increase the hole
concentration by two orders of magnitude compared to LT-MBE growth. Below 10 K, we observe negative magnetoresistance and anomalous Hall effect,
exhibiting the same hysteresis as the magnetization. In contrast, such effects are absent in samples with a smaller hole concentration. The
magnetic and magneto-transport properties can be qualitatively well explained within a picture of dopant segregation and the formation of bound
magnetic polarons.

\begin{acknowledgements}
Financial support from the Bundesministerium f\"{u}r Bildung und Forschung (FKZ13N10144) is gratefully acknowledged.
\end{acknowledgements}

\end{document}